\documentstyle[12pt,aasms]{article}
\tightenlines

\begin{document}

Accepted for publication in the Astrophysical Journal

\vskip 48pt

\title
{A Study of Absorption Features in the Three Micron Spectra of
Molecular Cloud Sources with H$_2$O Ice Bands}

\vskip 48pt

\author
{T. Y. Brooke$^ 1$}

\vskip 12pt

\affil
{Jet Propulsion Laboratory, M/S 169-237, 4800 Oak Grove Dr.,
Pasadena, CA  91109  USA}

\affil
{e-mail: tyb@scn5.jpl.nasa.gov}

\author
{K. Sellgren$^ {1}$}

\vskip 12pt

\affil
{Department of Astronomy, Ohio State University,
174 West 18th Av., Columbus, OH  43210  USA}

\affil
{e-mail: sellgren@payne.mps.ohio-state.edu}

\vskip 24pt

\author
{R. G. Smith}

\vskip 12pt

\affil
{Department of Physics, University College, University of New South Wales,
Australian Defence Force Academy, 2600 Canberra, Australia}

\affil
{e-mail: r-smith@adfa.edu.au}

\vskip 24pt

\altaffiltext{1}
{Visiting astronomer at Kitt Peak National Observatory}

\clearpage

\centerline
{\bf Abstract}

New 3.3--3.6~$\mu$m spectra were obtained of nine young stellar
objects embedded in molecular clouds.  An absorption feature at
$\sim$3.47~$\mu$m (2880~cm$^{-1}$) with FWHM$\sim$0.09 $\mu$m (80
cm$^{-1}$), first identified by Allamandola et al. (1992), was
definitively detected toward seven objects, and marginally in the
other two.  The feature is better correlated with H$_2$O ice than with
the silicate dust optical depth in the data obtained to date.
Assuming the feature is due to a C--H stretch absorption, the
abundance of the C--H bonds averaged along the lines of sight is
closely related to that of H$_2$O ice.  We interpret the correlation
with H$_2$O ice as indicating that the C--H bonds form together with
H$_2$O ice on grain surfaces in the molecular clouds, though other
formation mechanisms are not ruled out.  A second absorption feature
at 3.25 $\mu$m (3080 cm$^{-1}$) was detected toward NGC7538/IRS 1 and
S140/IRS 1; this feature was first detected in spectra of MonR2/IRS 3
(Sellgren, Smith, \& Brooke 1994; Sellgren et al. 1995).  There is as
yet insufficient data to tell whether this feature is better
correlated with H$_2$O ice or silicates.

\vskip 24pt

\keywords{infrared: general --- ISM: {dust,} extinction --- stars: pre-main
sequence}

\clearpage

\section
{Introduction}

\vskip 24pt

The 3.1 $\mu$m (3200 cm$^{-1}$) H$_2$O ice absorption band seen toward
molecular clouds has a long wavelength wing not accounted for by
simple models of grains covered only with H$_2$O ice (e.g. Smith,
Sellgren, \& Tokunaga 1989).  As the wing extends through the 3.2--3.6
$\mu$m (3100-2800 cm$^{-1}$) region characteristic of C--H stretch
vibrations, it is a logical place to search for the signatures of
solid organics.  Spectra of embedded young stellar objects in this
region have so far revealed the presence of solid CH$_3$OH at 3.54
$\mu$m (2825 cm$^{-1}$) and a 3.47 $\mu$m (2880 cm$^{-1}$) absorber
along the lines of sight (Grim et al.  1991; Allamandola et al. 1992;
Sellgren, Smith \& Brooke 1994).  Allamandola et al. suggested that
the 3.47 $\mu$m feature might be due to the C--H stretch absorption of
solo hydrogens attached to $sp^3$ bonded carbon clusters, the
``diamond''-like form of carbon.

New spectra from 3.3--3.6 $\mu$m (3000-2750 cm$^{-1}$) of nine young
stellar objects with H$_2$O ice features were obtained to determine
the prevalence and abundance of the 3.47 $\mu$m absorber in molecular
cloud dust.  Three sources were also observed from 3.0--3.3 $\mu$m
(3300-3000 cm$^{-1}$) to investigate a new absorption feature at 3.25
$\mu$m (3080 cm$^{-1}$) recently discovered toward MonR2/IRS 3
(Sellgren, Smith \& Brooke 1994; Sellgren et al. 1995).  This feature
may be due to aromatic hydrocarbons at low temperature.  The principal
question addressed is whether absorption features are correlated with
the H$_2$O ice or are due to components associated with the refractory
dust.

\goodbreak
\vskip 24pt

\section
{Observations}

\vskip 24pt

   Spectra were obtained at the Kitt Peak National Observatory 2.1-m
telescope on 14-19 Oct UT 1994.  The facility infrared cryogenic
spectrometer (CRSP) equipped with a 256$\times$256 InSb array was used
with a 300 lines/mm grating, which provided a spectral resolution
$\lambda$/$\Delta \lambda$~$\sim$~1300.  There were two pixels per
resolution element.  The wavelength calibration was obtained by
observing telluric OH emission lines in second order and is estimated
to be accurate to 5 \AA.

   The slit was $1.7^{\prime\prime} \times 156^{\prime\prime}$ with
the long axis oriented E--W.  The spatial resolution was 0.61
$^{\prime\prime}$/pixel.  Observations of each source were alternated
every 10--20 seconds with observations of blank sky 30-100$''$ away
from the source for background subtraction.  The blank sky positions
were chosen to minimize the contribution from any nebulosity or
additional nearby sources, using K$^{\prime}$ images of Hodapp (1994)
when available.  Five background columns on either side of the flux
peak on the array (3-6$^{\prime\prime}$ away) were fit for additional
background subtraction.

   All sources were observed at the peak of the 3.45 $\mu$m (2900
cm$^{-1}$) signal.  For those sources observed at 3.0--3.3 $\mu$m, the
telescope was first positioned to the 3.45 $\mu$m peak, since
surrounding nebulosity could cause the peak position to be a function
of wavelength in the strong H$_2$O ice absorption band.  Spectra from
3.0--3.3 $\mu$m were multiplied by a small correction factor to match
the flux in the region of overlap with the longer grating setting.
This correction was never more than 15 \%.  The flux differences were
likely the result of small centering and tracking errors.

   The stars used for atmospheric correction were also used for flux
calibration. The stars were assumed to be blackbodies.  The assumed
blackbody temperatures based on spectral type were from Lang (1992)
and are given in Table 1.  Blackbody functions were normalized to the
V magnitude fluxes.  The absolute calibrations may be uncertain to 20
\%.  Hydrogen absorption in the Pf $\delta$ line at 3.296 $\mu$m (3034
cm$^{-1}$) in the standard stars was not corrected for and most likely
contributes to the Pf $\delta$ emission features seen in some of the
objects.  Based on the spectral types of the standards, we estimate
that Pf $\delta$ absorption in the standards could contribute all or
part of the apparent Pf $\delta$ emission features in Elias 18 and NGC
7538/IRS 1, but not contribute significantly to the features in BN and
GL961E.  Some data points in the strongest telluric absorption
features were not well-corrected and were dropped.  Other gaps in the
spectra are due to occasional bad pixels.  Table 1 is a log of the
observations.  The spectra are shown in Fig. 1.

\vskip 24pt

\section
{Determination of Spectral Features}

\vskip 24pt

The intrinsic spectral shape of the combination of H$_2$O ice and the
long wavelength wing in molecular cloud sources has not been matched
either in theory or experimentally (cf. Smith, Sellgren \& Tokunaga
1989).  Therefore only an approximate local continuum can be defined.
Based on the width of the 3.47 $\mu$m and 3.25 $\mu$m features
identified in W3/IRS 5 (Allamandola et al. 1992) and Mon R2/IRS 3
(Sellgren, Smith \& Brooke 1994; Sellgren et al. 1995), the local
continuum was defined to be represented by points in the ranges
3.13-3.17 $\mu$m (where available), 3.33-3.37 $\mu$m, and points
longward of 3.61 $\mu$m.  The corresponding wavenumber ranges are
3190-3150 cm$^{-1}$, 3000-2970 cm$^{-1}$, and less than 2770
cm$^{-1}$.

   The continuum baselines were estimated using low order polynomial
fits to the logarithms of the flux densities of the continuum points.
Taking the logarithm enables spectra of the deepest ice band sources
to be fit better with low order polynomials, because of the natural
concavity of deep ice bands on linear scales.  We tested linear, 2nd,
and 3rd order polynomials.  For those sources observed at 3.0-3.3
$\mu$m, two separate polynomials of the same order joined together by
averaging in the central region of overlap were also tested.  We chose
as the best fit the one which minimized the deviations in the points
held to be continuum, subject to the fit passing over most of the
points held to contain a feature.  In each case, the best fit was a 3rd
order polynomial or the joining of two 3rd order polynomials.  In a
few cases where there were clearly no strong features, the continuum
was allowed to extend down to 3.58 $\mu$m (2790 cm$^{-1}$) to improve
the baseline at the longer wavelengths.  The adopted continua are
shown in Fig. 1.

   The resulting optical depths are shown in Fig. 2.  The 3.47 $\mu$m
and 3.25 $\mu$m features were fit with gaussians with the central
wavelength, peak depth, and full-width at half-maximum (FWHM) as free
parameters.  Uncertainties in these parameters were estimated from the
standard deviations of the results for fits using all of the baselines
that looked reasonable, i.e. that did not dip below the spectra.
Table 2 summarizes the results.  Individual objects are discussed
below.  There is no particular significance attached to the use of
gaussians; the purpose of the fitting is only to extract the best
estimates of the parameters of symmetric features from noisy data,
i.e. using all of the points available.

   For the weakest 3.47 $\mu$m features (S255/IRS 1, Elias 18) the
detections are marginal.  But for the other objects, the features
appear to be real, and the uncertainties simply reflect uncertainties
in the baseline.

It is important to note that the actual absorption profiles of the
species responsible for the features may extend further up into the
long wavelength wing. Since we fit only a local continuum, our
technique is sensitive only to the excess absorption at 3.47 $\mu$m or
3.25 $\mu$m above this continuum, and the derived optical depths may
be lower limits to the true contributions of the absorbers.

   Three objects (S140/IRS 1, W3/IRS 5, and NGC7538/IRS 9) were also
observed by Allamandola et al. (1992).  Our estimates of the 3.47
$\mu$m band depths differ somewhat from theirs as described below.  We
did not observe W33A, the other object studied by Allamandola et al.
However, we would interpret their 3.47 $\mu$m band depth as a lower
limit.  This is because there appears to be a substantial filling in
of the bottom of the 3.1 $\mu$m H$_2$O ice band in this object,
presumably by scattered light near the protostar.  This can be seen by
comparing the ratio of 3.47 to 3.35 $\mu$m optical depths of W33A,
0.84 (Willner et al. 1982) to those of the other ice-band sources for
which the ratio has been well-measured, 0.6-0.7 (Smith, Sellgren, \&
Tokunaga 1989).  A better estimate of the 3.47 $\mu$m band depth in
W33A will require a model of the 3 $\mu$m spectrum which includes
scattered light.

\vskip 24pt

\section
{Discussion of Individual Objects}

\vskip 24pt

\noindent
{\bf NGC7538/IRS 1}--This object appears to have both 3.47 and 3.25
$\mu$m features with peak optical depths $\tau\sim0.05$ and
$\tau\sim0.08$, respectively.  No simple continuum was found to fit
the entire range.  The adopted continuum consists of two third-order
polynomials averaged in the central continuum region.  Points near 3.3
$\mu$m containing Pf $\delta$ emission were deleted from the optical
depth plot.

\noindent
{\bf S140/IRS1}--This object has a 3.47 $\mu$m feature with peak
optical depth $\tau \sim 0.03$.  The shorter wavelength region suffers
from poor sky cancellation but is consistent with a 3.25 $\mu$m
feature with $\tau \sim 0.04$.  The spectrum in Fig. 1 is different
from that obtained by Allamandola et al. (1992), from which a 3.47
$\mu$m optical depth of 0.14 was estimated.  The main difference is at
wavelengths shortward of 3.45 $\mu$m, where the scatter in the
Allamandola et al. spectrum is high.  The fluxes in that spectrum are
also a factor $\sim4$ lower than in the present spectrum.  Both the
absolute flux and the spectral shape of the present spectrum are in
agreement with the earlier spectrum of Willner et al. (1982), however.

\noindent
{\bf BN}--This object has a 3.47 $\mu$m feature with $\tau \sim 0.03$.
The sharp feature at 3.10 $\mu$m and the broad dip near 3.2 $\mu$m are
due to the presence of crystalline H$_2$O ice (cf Smith, Sellgren \&
Tokunaga 1989).  Because of this structure, no simple continuum was
found to fit the entire range; only the long wavelength segments were
fit.  The hydrogen emission line in the Humphries series at 3.61
$\mu$m, H(20-6), was excluded from the continuum fit.  The other
emission lines indicated in Fig. 2 were excluded from the feature fit.
No firm statement can be made about a 3.25 $\mu$m feature; the
structure there is due to poor atmospheric cancellation.

\noindent
{\bf W3/IRS 5}--This object has a strong 3.47 $\mu$m feature with $\tau
\sim 0.13$.  It was also observed by Allamandola et al. (1992) who
derived a 3.47 $\mu$m band depth of 0.15 taking into account the
maximum possible contribution of solid CH$_3$OH.  As there is no clear
evidence for the spectral signature of solid CH$_3$OH at 3.54 $\mu$m
(2825 cm$^{-1}$) in the present spectrum, we attribute the entire band
to the 3.47 $\mu$m absorber.  Only the long wavelength range was
observed.

\noindent
{\bf NGC7538/IRS 9}--This object shows absorption by both solid
CH$_3$OH at 3.54 $\mu$m and the 3.47 $\mu$m absorber.  As shown by
Allamandola et al. (1992), solid CH$_3$OH will also contribute to the
absorption at shorter wavelengths.  For this reason, no attempt was
made to fit the entire feature with a single gaussian.  However, using
the solid CH$_3$OH absorption profile derived by Allamandola et al.
from laboratory spectra, we estimate peak optical depths of $\tau \sim
0.10$ for the 3.47 $\mu$m feature and $\tau \sim 0.08$ for CH$_3$OH at
3.54 $\mu$m with uncertainties of 0.02.  These values are about 60 \%
lower than those derived by Allamandola et al. due to the different
continuum adopted.

\noindent
{\bf GL 961E}--This object has a strong 3.47 $\mu$m feature with $\tau
\sim 0.07$.  The H(20-6) line was not included in the continuum fit and
the other emission lines indicated were not included in the feature fit.
Also excluded were points from 3.53-3.57 $\mu$m where there appears to
be additional absorption.  This could be due to solid CH$_3$OH (Grim
et al. 1991; Allamandola et al. 1992) with an optical depth at 3.54
$\mu$m of roughly 0.04, but the strong hydrogen emission lines make
this very difficult to confirm.

\noindent
{\bf MonR2/IRS 2}--This object has a 3.47 $\mu$m feature with $\tau
\sim 0.05$.  Two regions with apparent additional absorptions near
3.41 $\mu$m and 3.53 $\mu$m were excluded from the feature fit.  It is
not clear whether these features are real or due to incomplete
cancellation of telluric features which lie near these wavelengths.

\noindent
{\bf S255/IRS1}--This source appeared double at 3.3--3.6 $\mu$m.  The
sources were separated by 2.4$^{\prime\prime}$ E--W.  The eastern
source was 30 \% brighter and had a redder slope, presumably due to a
deeper H$_2$O ice band.  Its spectrum is used here.  It may have a
weak 3.47 $\mu$m feature with $\tau \sim 0.02$.  The western source
has at best only a weak 3.47 $\mu$m feature; the 2$\sigma$ upper limit
is $\tau \leq 0.03$.

\noindent
{\bf Elias 18}--Although the spectrum does not have high
signal-to-noise, this object appears to have a weak 3.47 $\mu$m
feature with $\tau \sim 0.02$.

\vskip 24pt

\goodbreak
\section
{Discussion}

\vskip 12pt

\subsection
{The 3.47 $\mu$m Feature}

\vskip 12pt

The 3.47 $\mu$m feature was detected in 7 of 9 young stellar objects
observed, and marginally detected in the remaining two (S255/IRS 1 and
Elias 18).  The absorber is clearly a widespread component towards
molecular cloud sources, as suggested by Allamandola et al.  (1992).
For the 10 young stellar objects observed here and by Sellgren, Smith,
\& Brooke (1994), the average central wavelength is $\lambda_0=3.469
\pm 0.002$ $\mu$m ($\nu_0=2883 \pm 2$ cm $^{-1}$) and the average FWHM
is $\Delta\lambda=0.092 \pm 0.005$ $\mu$m ($\Delta\nu=76 \pm 4$
cm$^{-1}$).  The uncertainties above are the 1$\sigma$ standard
deviations of the means of all the measurements.  The total spreads in
position and width are 0.023 $\mu$m (19 cm$^{-1}$) and 0.049 $\mu$m
(39 cm$^{-1}$), respectively.

To determine whether the 3.47 $\mu$m feature is due to a refractory
dust component, the peak optical depths are compared to the 9.7 $\mu$m
silicate dust absorption band depths in Fig. 3.  There is some
uncertainty in the silicate band depths due to the unknown amount of
intrinsic silicate emission in the sources.  All of the 9.7 $\mu$m
band depths except that of Elias 18 assume intrinsic silicate emission
in the source (see Willner et al. 1982).

There is not a strong correlation between the optical depths of the
3.47 $\mu$m feature and the silicate band (correlation coefficient
$r=0.55$).  Note that NGC 7538/IRS 9, with a strong 3.47 $\mu$m
feature, has about the same silicate band depth as S140/IRS 1 and
MonR2/ IRS 3, which have much weaker 3.47 $\mu$m features.  The lack
of correlation is in contrast to the C--H stretch absorption features
of aliphatic hydrocarbons seen in the diffuse interstellar medium
towards galactic center sources, which correlate with the silicates
(Sandford, Pendleton, \& Allamandola 1995).

There is a better correlation between the 3.47 $\mu$m absorber and
the 3.1 $\mu$m H$_2$O ice band ($r=0.89$) as shown in Fig. 4.
The best linear fit, exclusive of the upper limit, is
$$
\tau(3.47) = (0.035 \pm 0.003) \ \tau(3.1) - (0.014 \pm 0.006)   \eqno(1)
$$
which passes near the origin.  Thus the abundance of the 3.47
$\mu$m absorber is closely related to that of H$_2$O ice.

The optical depth of the 3.47 $\mu$m feature is also well-correlated
with the optical depth of the long wavelength absorption wing of the
H$_2$O ice band at 3.5 $\mu$m ($r=0.84$) as shown in Fig. 5.  The
optical depth of the long wavelength wing was estimated from low
resolution spectra by fitting blackbody curves to data at 2.5 and 3.8
$\mu$m (see Willner et al. 1982; Smith, Sellgren, \& Tokunaga 1989).
A linear fit gives
$$
\tau(3.47) = (0.17 \pm 0.02) \ \tau_{wing}(3.5) - (0.015 \pm 0.001).  \eqno(2)
$$
In both Eqs. 1 and 2, the uncertainties are the formal ones derived
from the measurement uncertainties; the true uncertainties can better
be gauged by noting the deviations of the points from the straight
lines.  Figs. 4 and 5 also demonstrate that the wing is fairly well
correlated with the peak optical depth in the ice band toward
protostars, a result which has been shown to hold for the Taurus dark
cloud medium (Smith, Sellgren, \& Brooke 1993).

The 3.47 $\mu$m feature may be simply structure on the long wavelength
wing, the overall profile of which is similar in many young stellar
objects as indicated in Fig. 6 of Smith, Sellgren, \& Tokunaga (1989).
The wing is believed to be due to overlapping absorptions of C--H
groups, though some kind of hydrate with H$_2$O has not been
completely ruled out.  Even if the absorption profile of the species
responsible for 3.47 $\mu$m feature extends into the wing, the
correlation with H$_2$O ice should still hold, since the wing as a
whole correlates fairly well with H$_2$O ice.  The correlation breaks
down only in the worst case scenario: a) the 3.47 $\mu$m feature
extends into the wing; b) it extends by different amounts relative to
the rest of the wing in different sources; and c) some other absorber
fills in the rest of the wing so as to mimic a correlation with H$_2$O
ice.  We consider this an unlikely possibility, though it can't be
ruled out.

{}From the peak wavelength of the 3.47 $\mu$m feature, Allamandola et
al.  (1992), proposed that single hydrogens attached to $sp^3$ bonded
carbon clusters might be responsible for the feature.  Assuming the
feature to be due to C--H bonds, Fig. 4 indicates that averaged along
the lines of sight, the ratio of abundances of the C--H bonds and
H$_2$O ice is roughly similar.

The lack of a correlation with silicates suggests that the C--H bonds
are not associated with the refractory dust cores.  Some other
plausible origins for the C--H bonds are: direct condensation of
hydrocarbons from the gas, ultraviolet or thermal processing of
carbon-containing ices as demonstrated in the laboratory (e.g.
Allamandola, Sandford, \& Valero 1988; Schutte, Allamandola, \&
Sandford 1993), or surface reactions on the grains.

One fact worth noting is that the 3.47 $\mu$m feature is nearly as
strong relative to H$_2$O ice in NGC7538/IRS 9 as it is in W3/IRS 5,
though NGC7538/IRS 9 has enhanced ratios of solid CO and CH$_3$OH
relative to H$_2$O compared to W3/IRS 5 (Allamandola et al. 1992).
Thus the abundance of the C--H bonds does not appear to correlate with
solid CO, which may primarily condense from the gas (d'Hendecourt,
Allamandola, \& Greenberg 1985; Brown and Charnley 1990).  Nor does
the growth of the 3.47 $\mu$m feature appear to come at the expense of
CO (or CH$_3$OH) by ultraviolet processing (assuming that both clouds
began with the same complement of ices) as has recently been suggested
for the 2166 cm$^{-1}$ ``X--CN'' band (Tegler et al. 1995).  This does
not strictly rule out an ultraviolet or thermal processing origin for
the 3.47 $\mu$m feature, though it suggests that different initial
abundances, parents or processes would have to occur in different
clouds, or in different regimes along the lines of sight.

Theoretical calculations suggest that H$_2$O ice in molecular clouds
forms primarily by surface reactions on grains rather than
condensation from the gas (Jones \& Williams 1984; d'Hendecourt,
Allamandola, \& Greenberg 1985; Brown \& Charnley 1990).  If we assume
that the C--H bonds also form by surface reactions, then we are led to
interpret the correlation of the 3.47 $\mu$m feature with H$_2$O ice
as indicating that the C--H bonds form together with H$_2$O ice on
grain surfaces in molecular clouds.

Duley and Williams (1995) have recently proposed that H$_2$O ice and
hydrogenated amorphous carbon (proposed to account for the long
wavelength wing) both form by surface reactions on the grains with the
formation of $sp^3$ bonded carbon being necessary for the retention of
an H$_2$O ice mantle.  The correlation of the 3.47 $\mu$m feature with
H$_2$O ice provides support for such a scenario, if in fact the
feature is due to C--H groups on carbon clusters.  However, the
observed correlation does not necessarily require the C--H bonds and
H$_2$O ice to form on the same grains, or for the C--H bonds to form
first as Duley and Williams propose.

Under the assumption that the 3.47 $\mu$m feature is due to C--H groups
on carbon, the required column densities (Table 3) were estimated from
$N=\tau\Delta\nu/A$, where $\tau$ is the peak optical depth,
$\Delta\nu$ the FWHM in cm$^{-1}$, and $A$ the integrated absorbance, taken to
be $4.0 \times 10^{-18}$ cm per C--H bond (Allamandola et al.
1992).  With $A=2.0 \times 10^{-16}$ cm for the 3.1 $\mu$m
H$_2$O ice band (Allamandola, Sandford, \& Valero 1988), the linear
relation of Fig. 4 gives $N_{C-H}/N_{H_2O}\approx0.42$ or roughly one
C--H bond for every two H$_2$O molecules.

It is difficult to say with certainty at this time whether the 3.47
$\mu$m is the signature of solo C--H bonds on $sp^3$ bonded carbon
clusters as proposed by Allamandola et al. (1992).  There has been
much work recently on characterizing amorphous and diamond-like carbon
films in the laboratory.  Assignments for the monohydride stretch in
these films vary from 3.42 $\mu$m (2920 cm$^{-1}$) to 3.53 $\mu$m
(2830 cm$^{-1}$) (Dischler, Bubenzer, \& Koidl 1983; Chin et al.
1992).  These bracket the observed feature, so the Allamandola et al.
identification is plausible.  Relevant experiments at low temperatures
would be useful to test whether material can be created with a band at
the correct position and width under interstellar conditions.

\vskip 12pt

\subsection
{The 3.25 $\mu$m Feature}

\vskip 12pt

A 3.25 $\mu$m feature similar to that identified in Mon R2/IRS 3
(Sellgren, Smith, \& Brooke 1994; Sellgren et al. 1995) was detected
in NGC 7538/IRS1 and S140/IRS 1.  The telluric atmosphere contains
strong features at this wavelength; however the present observations
add to the evidence that there is a real absorption feature here.  The
wavelength of the 3.25 $\mu$m feature is near the range expected for
hydrogens attached to aromatic hydrocarbons at low temperatures.  This
feature is discussed further by Sellgren et al. (1995).

The 3.25/3.47 optical depth ratios are roughly comparable in all three
objects.  However, there is as yet insufficient data to tell whether
the 3.25 $\mu$m feature is better coupled to H$_2$O ice or silicate
dust.  Further searches for the 3.25 $\mu$m feature and determination
of possible correlations are planned.

\vskip 24pt

\goodbreak
\section
{Conclusions}

\vskip 12pt

  A distinct absorption feature at $\sim$3.47~$\mu$m (2880~cm$^{-1}$)
with FWHM$\sim$0.09 $\mu$m (80 cm$^{-1}$), first identified by
Allamandola et al. (1992), is extremely common toward young stellar
objects with H$_2$O ice bands.  The feature may be structure on the
broad long wavelength wing of the ice band that remains after the
absorption by pure H$_2$O ice has been removed.  The 3.47 $\mu$m
feature, like the long wavelength wing, is better correlated with
H$_2$O ice than with the silicate dust optical depth in the data
obtained to date.  If the feature is due to a C--H stretch absorption,
such as the solo hydrogens on $sp^3$ bonded carbon clusters suggested
by Allamandola et al., then we interpret the correlation with H$_2$O
as indicating that the C--H bonds form together with H$_2$O ice on
grains in the molecular cloud, with roughly one C--H bond forming for
every two H$_2$O molecules, though other formation mechanisms are not
ruled out.

There are now three detections of a second distinct absorption feature
at 3.25 $\mu$m (3080 cm$^{-1}$) toward young stellar objects; this
feature may be due to aromatic hydrocarbons at low temperature along
the lines of sight.  There is as yet insufficient data to determine
whether this feature is correlated with H$_2$O ice or not.

\vskip 24pt

\acknowledgments

\vskip 12pt

We thank R. Joyce (NOAO) for assistance with the instrument and T.
Mailloux (OSU) for help with the observations.  K. Hodapp (U. Hawaii)
provided digital images of some of the objects from his K$^{\prime}$
survey.

\vskip 24pt

\clearpage
\begin{table}
\begin{center}
\begin{tabular}{lcccccc}
\multicolumn{7}{c}{{\bf Table 1: Log of Observations}}\\[12pt]

Object&UT&Range&t$^a$&Standard&Spectral&${T_{bb}}^b$ \\
&(Oct 1994)&($\mu$m)&(sec)&Star&Type&(K) \\[12pt]

W3/IRS5&14.3&3.29-3.63&120&BS 1035&B9Ia&10255\\
Elias 18&14.5&3.29-3.63&360&BS 1497&B3V&18700\\
BN&19.5&3.29-3.63&64&BS 1673&F2V&6890\\
  &19.5&3.02-3.38&96&&&\\
Mon R2/IRS2&14.5&3.29-3.63&360&BS 1931&O9.5V&32000\\
S255/IRS1&19.5&3.29-3.63&128&BS 2484&F5V&6440\\
GL961E&17.5&3.29-3.63&320&BS 1412&A7III&7650\\
S140/IRS1&19.2&3.29-3.63&240&$\delta$ Cep&F5Iab&6000\\
  &18.3&3.02-3.38&320&&&\\
NGC 7538/IRS1&18.2&3.29-3.63&320&$\beta$ Cas&F2IV&6870\\
  &17.2&3.02-3.38&256&&&\\
NGC 7538/IRS 9&18.4&3.29-3.63&240&$\beta$ Cas&F2IV&6870\\

\end{tabular}
\end{center}
\end{table}

Notes to Table 1:

{$^a$} Integration time

{$^b$} Assumed blackbody temperature of standard star.

\clearpage
\begin{table}
\begin{center}
\begin{tabular}{cccccc}
\multicolumn{6}{c}{{\bf Table 2: Absorption Feature Parameters$^{ab}$}}\\[12pt]
Object&${\lambda_0}$&${\nu_0}$&$\Delta\lambda$&$\Delta\nu$&$\tau$\\[12pt]
&($\mu$m)&(cm$^{-1}$)&($\mu$m)&(cm$^{-1}$)&\\[12pt]
\multicolumn{6}{c}{3.47 $\mu$m (2880 cm$^{-1}$) feature}\\[12pt]
W3/IRS5&3.461&2889&0.114&95&0.133\\
       &(0.003)&(3)&(0.010)&(8)&(0.019)\\
Elias 18&3.469&2883&0.117&98&0.022\\
       &(0.008)&(7)&(0.066)&(54)&(0.007)\\
BN&3.459&2891&0.093&78&0.034\\
       &(0.003)&(3)&0.021)&(18)&(0.010)\\
Mon R2/IRS2&3.476&2877&0.095&79&0.046\\
       &(0.004)&(3)&(0.014)&(12)&(0.009)\\
S255/IRS1&3.467&2885&0.068&56&0.024\\
       &(0.006)&(5)&(0.028)&(24)&(0.018)\\
GL961E&3.463&2888&0.099&82&0.068\\
       &(0.003)&(3)&(0.006)&(5)&(0.004)\\
S140/IRS1&3.473&2879&0.072&60&0.027\\
       &(0.007)&(6)&(0.032)&(26)&(0.007)\\
NGC 7538/IRS1&3.464&2887&0.092&77&0.052\\
       &(0.004)&(3)&(0.019)&(16)&(0.014)\\
NGC 7538/IRS 9&3.48&2880&0.08&70&0.10\\
       &(0.01)&(10)&(0.02)&(16)&(0.02)\\

\end{tabular}
\end{center}
\end{table}

Notes to Table 2:

{$^a$} Central wavelengths (frequencies), full widths at half maximum, and
peak optical depths of absorption features from gaussian fits.

{$^b$} Entries in parentheses are 1$\sigma$ uncertainties obtained
from standard deviations of results using several different baselines
(see text).

\clearpage
\begin{table}
\begin{center}
\begin{tabular}{cccccc}
\multicolumn{6}{c}{{\bf Table 2--continued}}\\[12pt]
Object&${\lambda_0}$&${\nu_0}$&$\Delta\lambda$&$\Delta\nu$&$\tau$\\[12pt]
&($\mu$m)&(cm$^{-1}$)&($\mu$m)&(cm$^{-1}$)&\\[12pt]
\multicolumn{6}{c}{3.25 $\mu$m (3080 cm$^{-1}$) feature}\\[12pt]
S140/IRS1&3.255&3072&0.073&69&0.036\\
       &(0.004)&(4)&(0.005)&(5)&(0.007)\\
NGC 7538/IRS1&3.263&3065&0.091&85&0.078\\
       &(0.008)&(8)&(0.006)&(6)&(0.013)\\

\end{tabular}
\end{center}
\end{table}

\clearpage
\begin{table}
\begin{center}
\begin{tabular}{ccccccc}
\multicolumn{7}{c}{{\bf Table 3: Dust Feature Optical Depths}}\\[12pt]

Object&$\tau$(9.7)&$\tau$(3.1)&$\tau_{wing}$(3.5)&$\tau$(3.47)&$\tau$(3.25)&N$_{C-H}$(3.47)\\[12pt]
&&&&&&(10$^{18}$ cm$^{-2}$)\\
&&&&&&\\
W3/IRS5&7.64$^a$&3.48$^c$&0.71$^c$&0.133&--&3.2\\
Elias 18&0.43$^b$&0.80$^b$&0.13$^2$&0.022&--&0.54\\
BN&3.28$^a$&1.78$^c$&0.22$^c$&0.034&--&0.66\\
Mon R2/IRS2&--&2.54$^c$&0.46$^c$&0.046&--&0.91\\
Mon R2/IRS3&4.30$^a$&1.14$^c$&0.39$^c$&0.036$^d$&0.049$^d$&0.68\\
S255/IRS1&5.11$^a$&1.48$^c$&0.26$^c$&0.024&--&0.34\\
GL961E&2.11$^a$&2.46$^c$&0.36$^c$&0.068&--&1.4\\
W33A&7.84$^a$&$>$5.4$^a$&2.53$^a$&$>$0.15$^1$&--&$>$3.0\\
S140/IRS1&3.97$^a$&1.28$^a$&0.46$^a$&0.027&0.036&0.40\\
NGC 7538/IRS1&6.38$^a$&1.29$^a$&0.42$^a$&0.052&0.078&1.0\\
NGC 7538/IRS 9&4.46$^a$&3.28$^a$&0.64$^a$&0.10&--&1.8\\

\end{tabular}
\end{center}
\end{table}

Notes to Table 3: The optical depths given in the table are for
silicates at 9.7 $\mu$m, $\tau$(9.7); H$_2$O ice at 3.08 $\mu$m,
$\tau$(3.1); the long wavelength wing of the ice band at 3.47 $\mu$m,
$\tau_{wing}$(3.5); and features at 3.47 and 3.25 $\mu$m.
N$_{C-H}$(3.47) is the column density of C-H bonds required to give
the 3.47 $\mu$m feature assuming an absorbance A=$4.0 \times 10^{-18}$
cm per C--H bond (see text).

References---
{\it (a)} Willner et al. 1982;
{\it (b)} Whittet et al. 1988.
{\it (c)} Smith, Sellgren, \& Tokunaga 1989;
{\it (d)} Sellgren, Smith \& Brooke 1994;

Notes---

(1) Optical depth from Allamandola et al. (1992)
interpreted as an lower limit.

(2) Estimated from Whittet et al. 1988.

\clearpage
\centerline
{\bf Figure Captions}

\vskip 12pt

{\bf Figure 1(a-c)}--- Spectra with resolution $\lambda$/$\Delta \lambda
\sim 1300$.  Error bars are $\pm$1-$\sigma$.  Solid lines are
polynomial fits to the observations in local continuum regions (see
text).

{\bf Figure 2(a-c)}---Optical depths from the continuum fits in Fig.
1.  Solid lines are gaussian fits to features.  Hydrogen emission
lines marked in BN and GL961E were excluded from the fits.

{\bf Figure 3}---3.47 $\mu$m feature optical depths vs. silicate
optical depths.  Error bars correspond to 1$\sigma$ uncertainties
obtained from the standard deviations of fits using several different
baselines (see text).

{\bf Figure 4}---3.47 $\mu$m feature optical depths vs. H$_2$O ice
optical depths.  Solid line is best linear fit.

{\bf Figure 5}---3.47 $\mu$m feature optical depths vs. long
wavelength wing optical depths.  Solid line is best linear fit.
Refer to Fig. 4 for identification of points.

\pagestyle{empty}
\clearpage

\begin{figure}
\plotfiddle{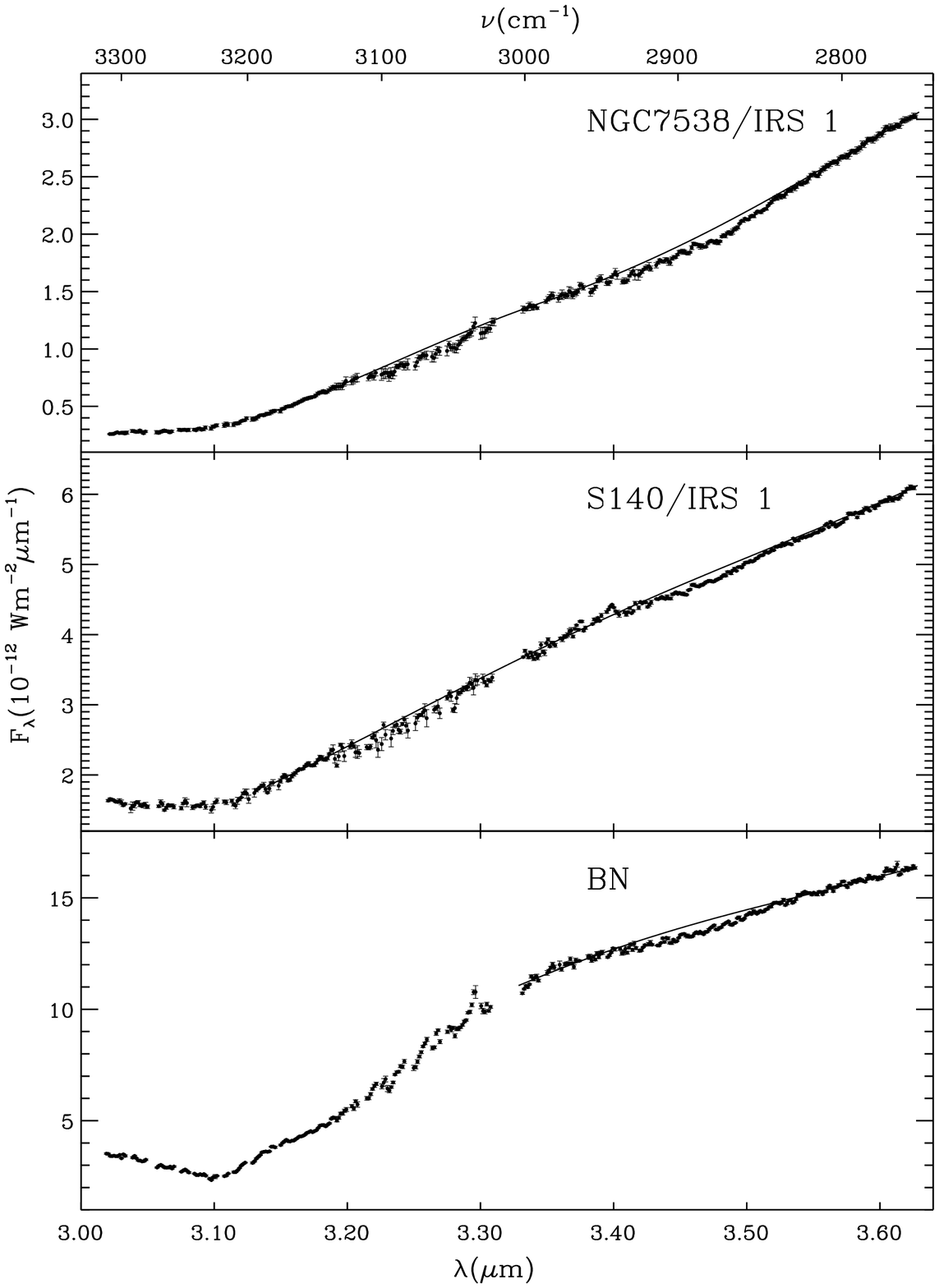}{9.0in}{0}{100}{100}{-324}{0}
\end{figure}

\begin{figure}
\plotfiddle{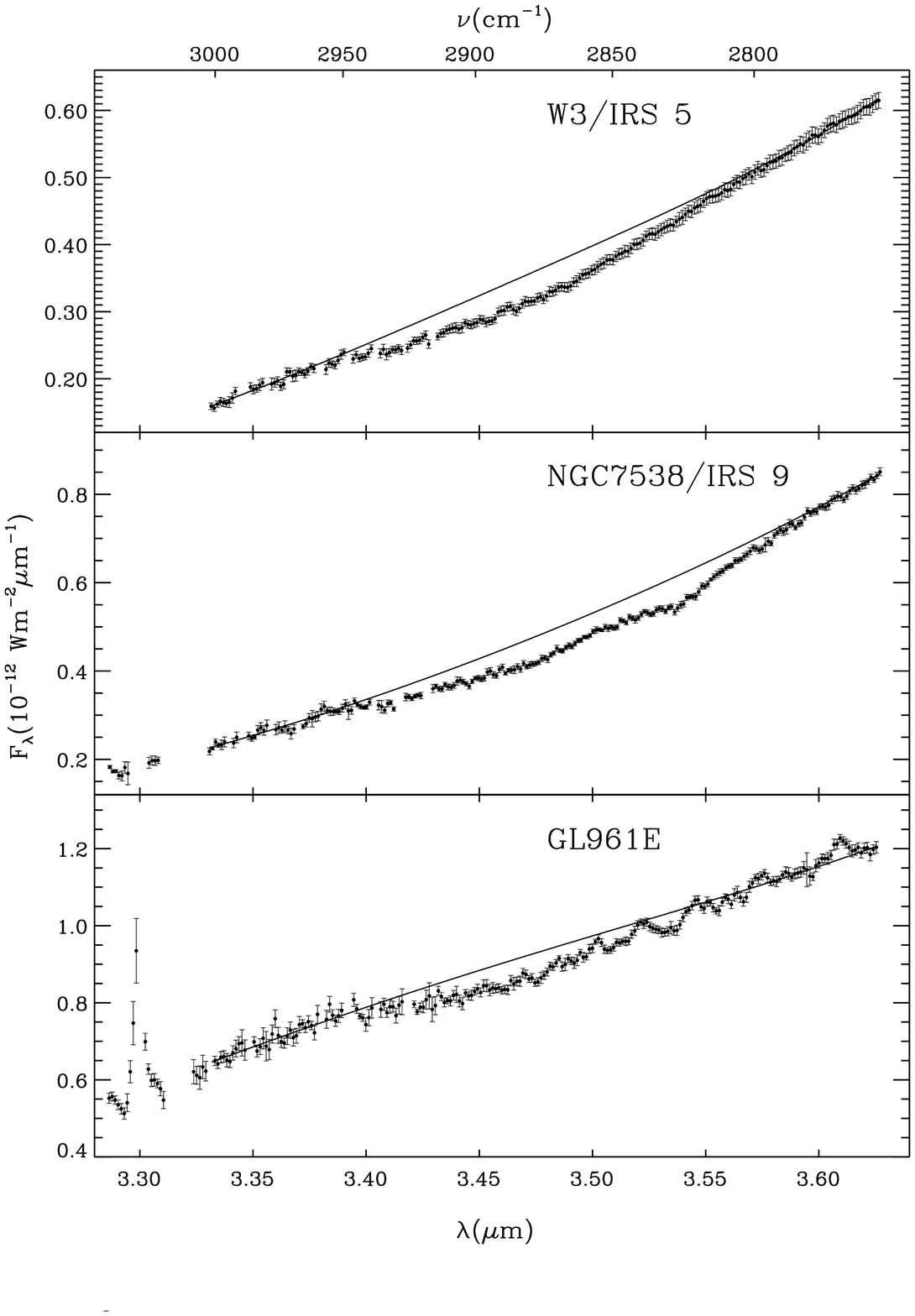}{9.0in}{0}{100}{100}{-324}{0}
\end{figure}

\begin{figure}
\plotfiddle{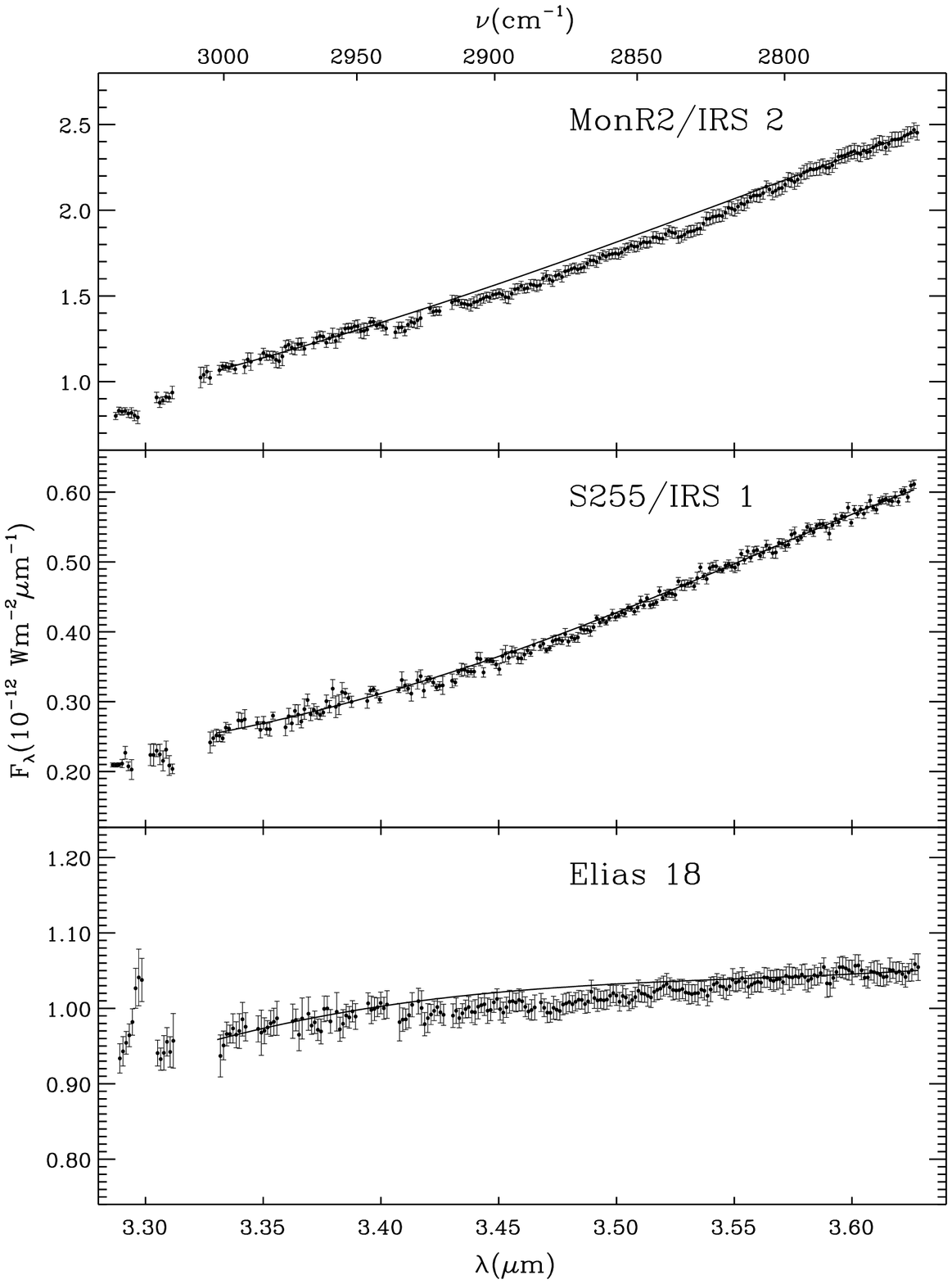}{9.0in}{0}{100}{100}{-324}{0}
\end{figure}

\begin{figure}
\plotfiddle{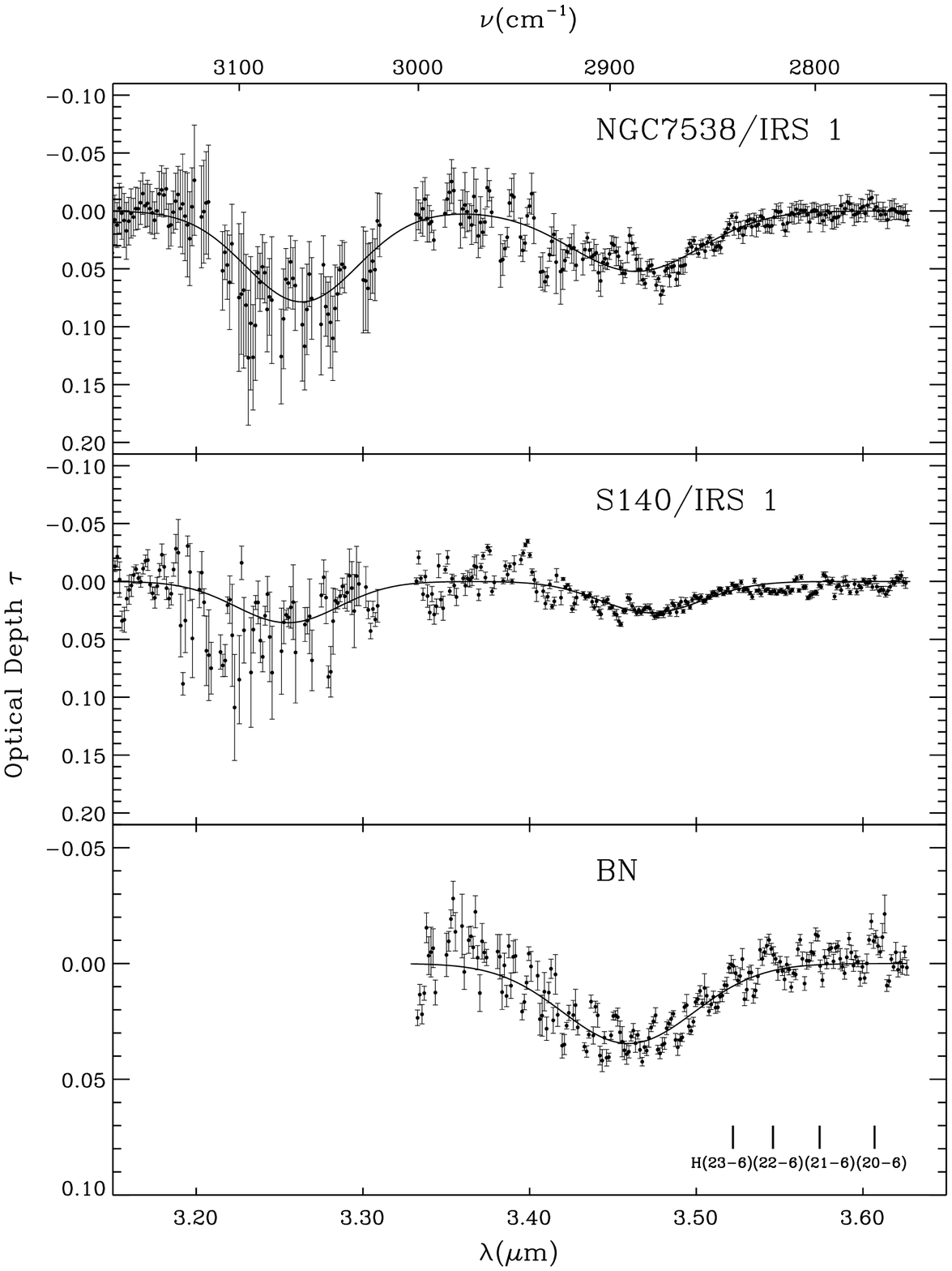}{9.0in}{0}{100}{100}{-324}{0}
\end{figure}

\begin{figure}
\plotfiddle{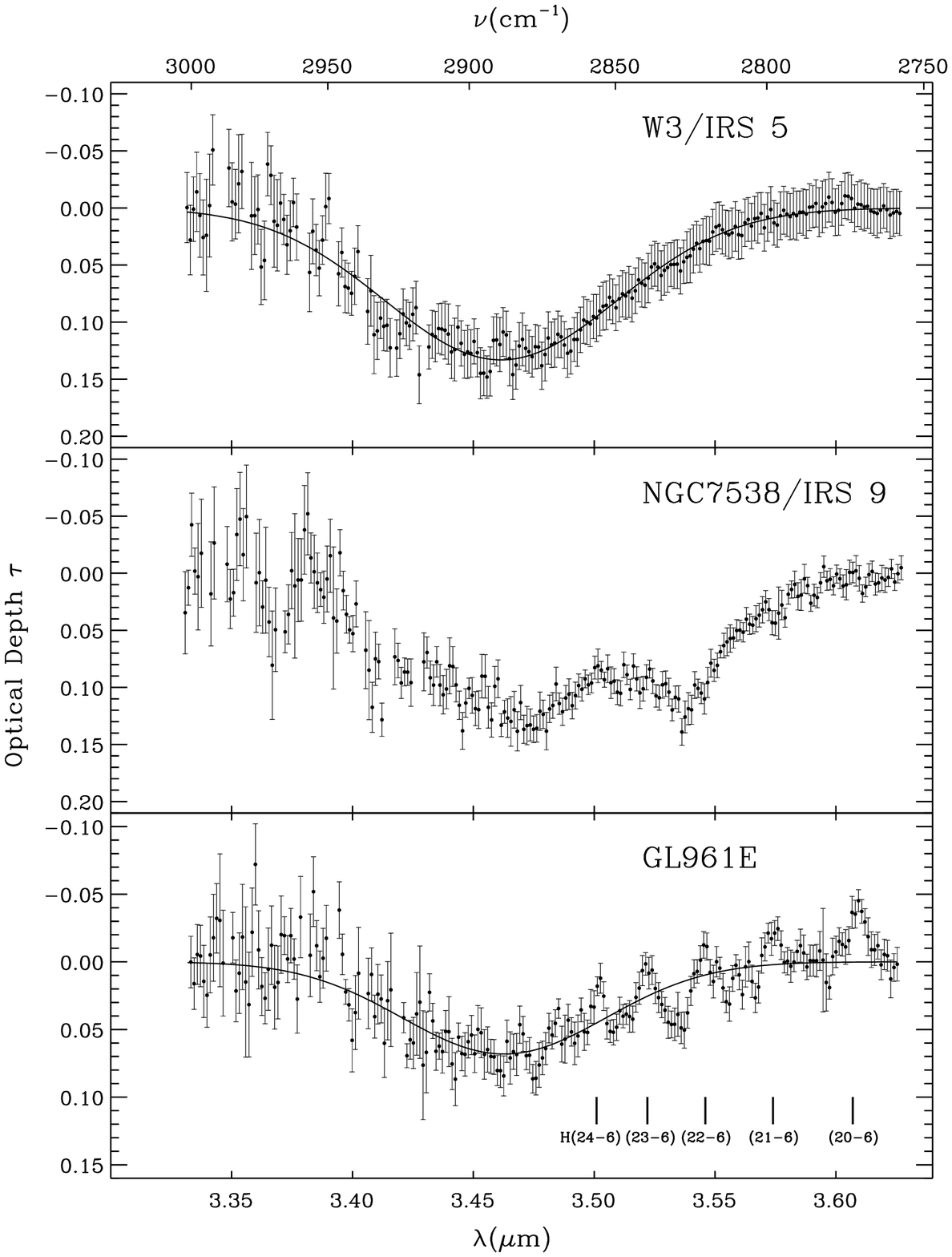}{9.0in}{0}{100}{100}{-324}{0}
\end{figure}

\begin{figure}
\plotfiddle{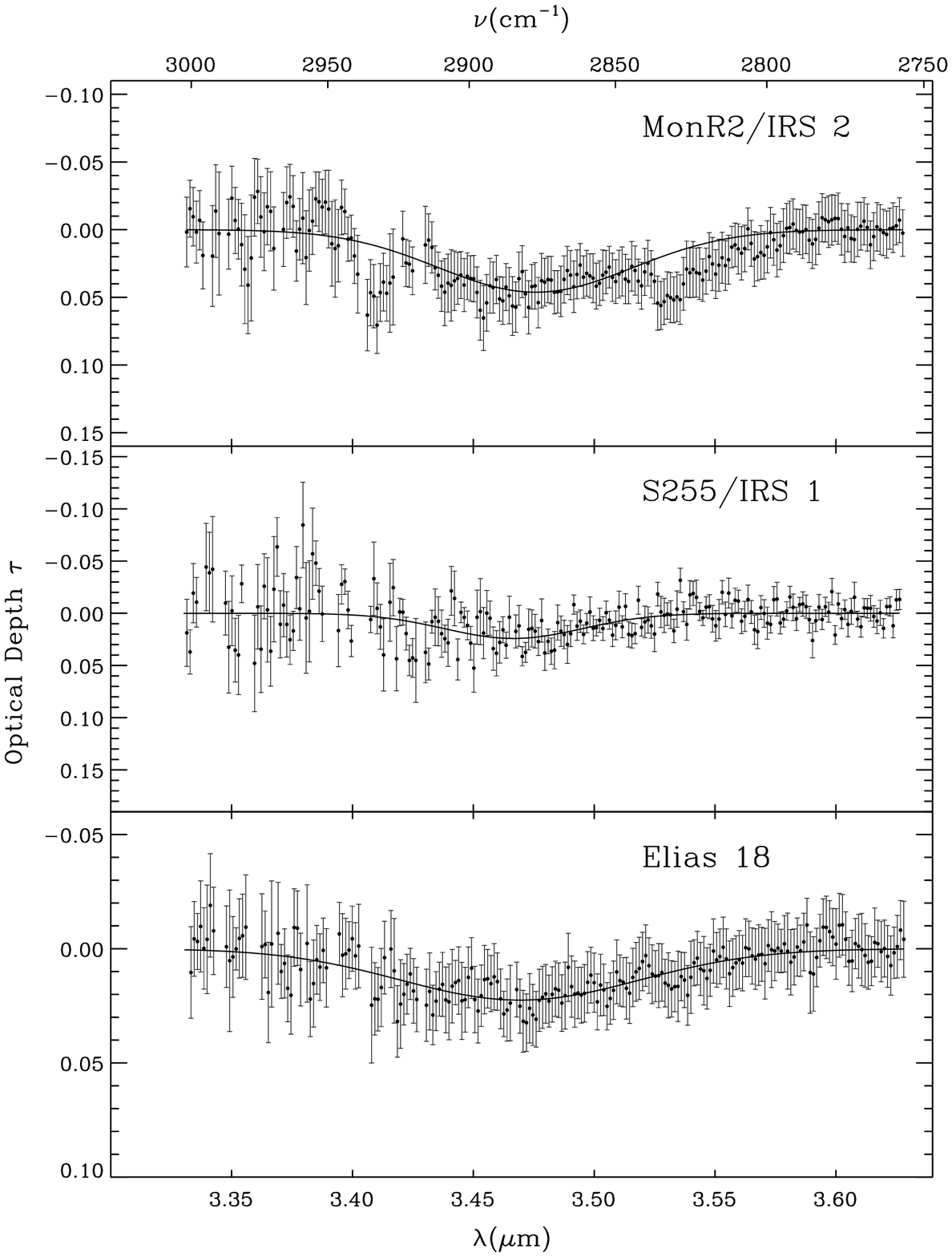}{9.0in}{0}{100}{100}{-324}{0}
\end{figure}

\begin{figure}
\plotfiddle{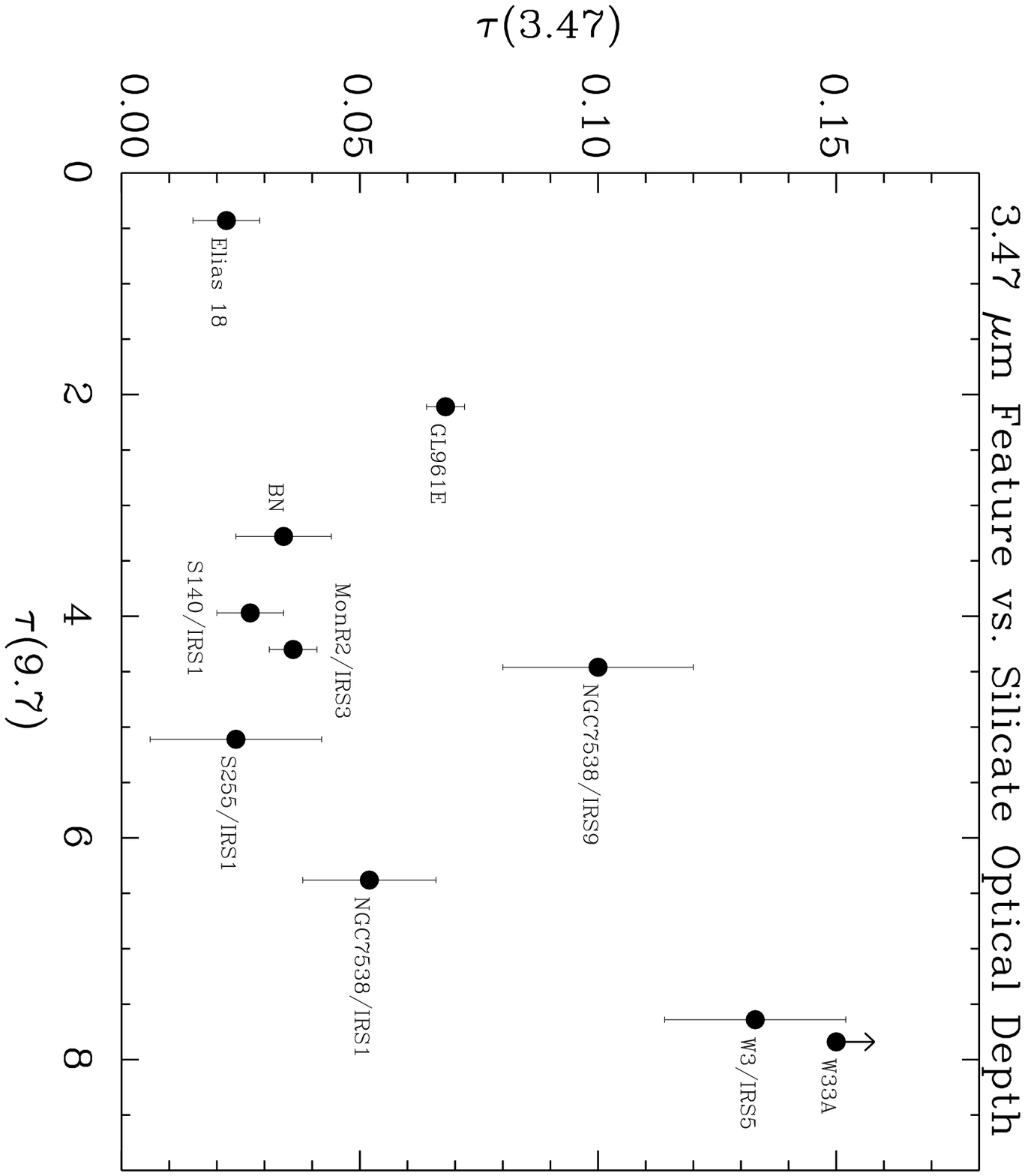}{9.0in}{0}{100}{100}{-324}{0}
\end{figure}

\begin{figure}
\plotfiddle{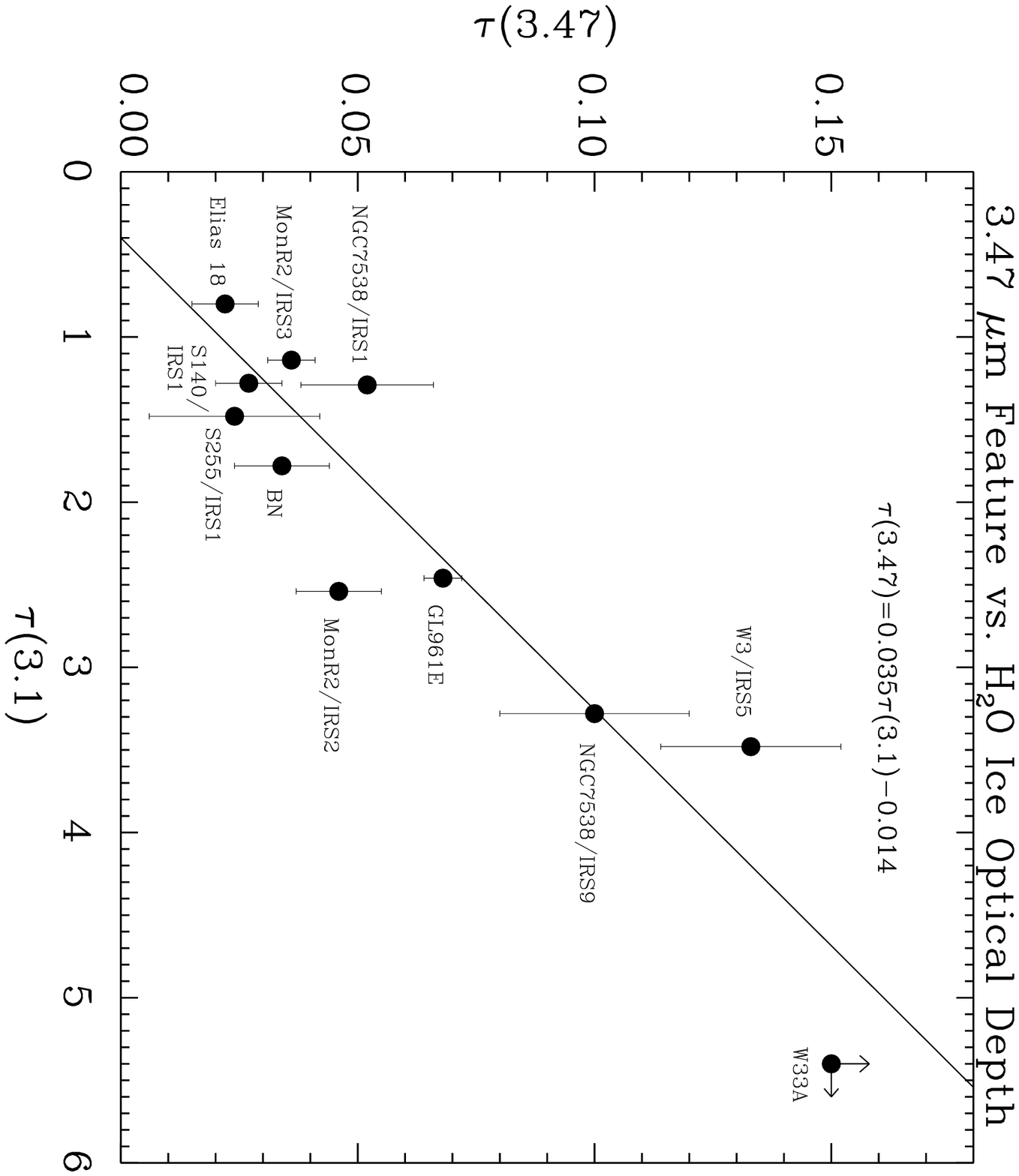}{9.0in}{0}{100}{100}{-324}{0}
\end{figure}

\begin{figure}
\plotfiddle{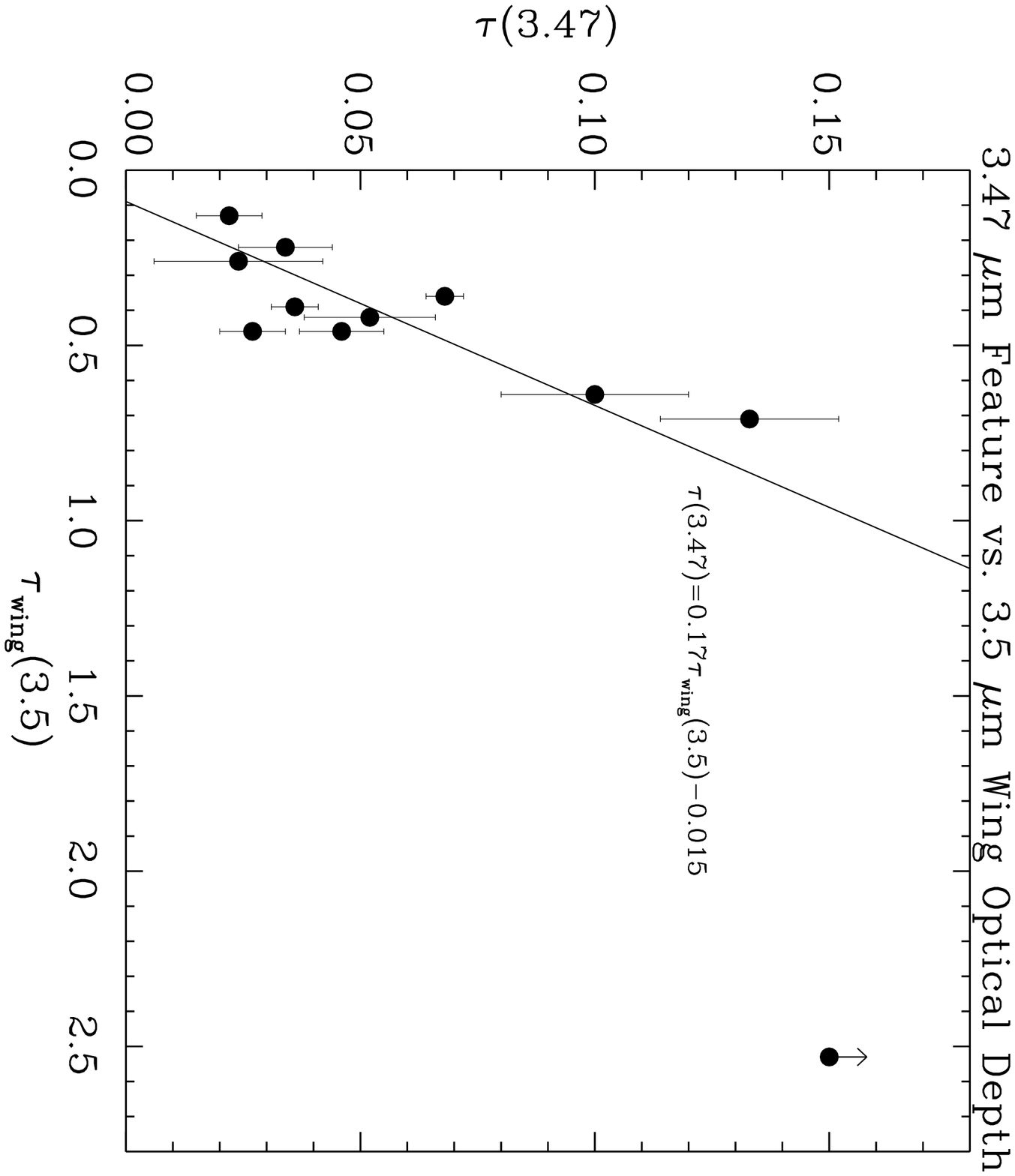}{9.0in}{0}{100}{100}{-324}{0}
\end{figure}

\end{document}